\documentclass{amsart}
\usepackage{amssymb}
\usepackage{amsfonts}
\usepackage{amsmath}
\usepackage[legalpaper,bookmarks=true,colorlinks=true,linkcolor=blue,citecolor=blue]{hyperref}
\usepackage{graphicx}%
\usepackage{fancyhdr}
\usepackage{color}
\usepackage[mathlines]{lineno}
\usepackage{lscape}
\usepackage{epsfig}
\usepackage{natbib}
\usepackage{geometry}
\usepackage{tgbonum}
\usepackage[]{listings}

\fontfamily{qcr}\selectfont

\theoremstyle{plain}

\numberwithin{equation}{section}

\newcommand{\Bin}{\bigskip \noindent}

\newcommand{\Ni}{\noindent}

\begin{document}
\Large
\title[A Supervised  Hybrid Statistical Catch-Up System]{A Supervised  Hybrid Statistical Catch-Up System Built on GABECE Gambian Data}

\author{Tagbo Innocent Aroh}
\author{Ousman Saine}
\author{Soumaila Demb\'el\'e}
\author{Gane Samb Lo}

\begin{abstract} In this paper we want to find a statistical rule that assigns a passing or failing grade to students who undertook at least three exams out of four in a national exam, instead of completely dismissing them students. While it is cruel to declare them as failing, especially if the reason for their absence it not intentional, they should have demonstrated enough merit in the three exams taken to deserve a chance to be declared passing. We use a special classification method and nearest neighbors methods based on the average grade and on the most modal grade to build a statistical rule in a supervised learning process. The study is built on the national GABECE educational data which is a considerable data covering seven years and all the six regions of the Gambia.\\

\Ni Tagbo Innocent Aroh $^{\dag}$\\
Department Of Mathematics \& Statistics\\
Auburn University, AL, USA\\
tia0002@auburn.edu, arohinnocent@gmail.com

\noindent Ousman Saine $^{\dag\dag}$\\
University of The Gambia, Banjul, The Gambia\\
\& Policy Planning, Analyses, research and Budgeting Directorate,\\
Ministry of Basic and secondary Education, Willy Thorpe Palace,\\
Banjul, The Gambia\\
ousmansaine@edugambia.gm\\

\noindent Soumaila DEMBELE $^{\dag\dag\dag}$\\
Université des Sciences Sociale et de Gestion de Bamako ( USSGB)\\
Faculté des Sciences Économiques et de Gestion (FSEG)\\
Email: soumidemlpot@gmail.com\\

\noindent Gane Samb Lo $^{\dag\dag\dag\dag}$.\\
LERSTAD, Gaston Berger University, Saint-Louis, S\'en\'egal (main affiliation).\newline
LSTA, Pierre and Marie Curie University, Paris VI, France.\newline
AUST - African University of Sciences and Technology, Abuja, Nigeria\\
gane-samb.lo@edu.ugb.sn, gslo@aust.edu.ng, ganesamblo@ganesamblo.net\\
Permanent address : 1178 Evanston Dr NW T3P 0J9,Calgary, Alberta, Canada.\\

\noindent\textbf{Keywords}. GABECE Gambian data; missing data; estimation of missing data; catch-up system; classification; K-Mean methods; k-NN; nearest neighbors techniques; R-codes.\\
\textbf{AMS 2010 Mathematics Subject Classification:} 60-07;\\

\end{abstract}
\maketitle

\newpage
\section{Introduction} \label{sec1}

\noindent The West African Examinations Council was established in 1952. This Council has the responsibility of ascertaining examinations required in the public interest in the English-speaking West African countries, comparable to those of equivalent examining authorities internationally, and to develop syllabuses, arranges and administers examinations and awards certificates. It also conducts a number international examinations.\\

\noindent The data we are going to use in this document is from the above aforementioned body. The data covers the period from the 2012 academic year to 2017. The mode of the GABECE examination is that each student is required to choose 7 subjects at least and 9 subjects at most. Out of the subjects taken, four (4) are compulsory and are called core subjects, which are English (ENG), Mathematics (MATH), Sciences (SCIENCES)  and Social and Environmental Studies (SES).\\

\noindent Individual performance grants the student admission to Senior Secondary Education. The grading system in this result is categorized as follows : Credit 
(grades from 1 to 6), Pass (7 to 8), Fail (9). The result of a student is determined by the grades of the four core subjects plus the two best grades from the student's choice. The best result a student could have is the aggregated grade from 6 (six), one in all the core subjects, plus one in the two choices of the students and the worst result is aggregated to 54 with failure  in all subjects with grade 9.\\

\Ni In the introductory paper on \textit{GABECE} (\cite{saine2020}) where database has been widely described, full notation related to the available variables are posted. So, we refer the reader to the aforementioned paper for a more general description.\\

\noindent Our study focuses on the core subjects which are English, Mathematics, Science and Social and Environmental Studies. So many times, it happen that one student is successful, some times with credits in three exams among four majors and fail unwillingly to be present at the fourth. Such situations may arise in case of illness, delays due to transportation, late wake up, sudden discomfort in the classroom, etc. If situations like these can be certified, what can be done to save these students instead of throwing them as failing or treating them as weak students?\\

\noindent One may argue that such students deserve to have a second chance by sitting for an organized second sessions. But organizing these sessions for all students in the described case, for all the four disciplines over the whole country may be difficult to implement. Based on such difficulties, it is natural to wonder if statistical rules can be built that allows granting a grade or an overall pass grade to such students without violating the merit-base system.\\

\noindent Covariates are the gender, the area and the year. So building models may be done for different levels of them : modeling for a specific gender, or an area or a year, or a combination of them.\\

\noindent We will see later that we have one hundred and eighty four (184) of students who took the three first exams and missed the fourth one, amongst them only four (4) passed the first three exams that they took, over a period of five years (from 2012 to 2017) for all regions. The situation happened three times in 2015 and the concerned regions were 2, 3 and 1, and one time in 2017 in region 1. Table \ref{tab1} shows those cases.\\

\begin{table}
\centering
\begin{tabular}{lllllll}
\hline \hline
Case  	& Year 	&  Region &  English 	& Maths  & Sciences 	& SES	\\
\hline \hline
 77594  & 2015  &  2      &   8       	& 8      & 8 		& -1    \\
77833   & 2015  &  3     &   8      	& 8      & 8      &  -1  	\\
80183   & 2015  &  1      &   4      	& 6      & 7      &  -1	\\
122915	& 2017  &  1      &   1       	& 7      & 7    & -1\\
\hline \hline
\end{tabular}
\caption{Cases passing English, Maths, Sciences grades and absent in the SES exam}
\label{tab1}
\end{table}

\Bin Finally, we succeeded in building a model with the following characteristics : fairness and usefulness. We have been able to apply to the four cases who missed the fourth exam  having passed the other three exams. A similar study is still to be done for another missed exam.\\

\noindent The rest of the paper organized as follows.\\

\noindent (a) We present the data in section \ref{sec2}, by describing the variables, their yearly, gender and region versions and provide the R codes to create them. This is justified by our wish to see other countries / institutions with similar challenges to be able to extend the research and the methods to their data. We finish that section by some statistical facts we drew from a direct unsupervised learning process.\\ 

\noindent (c) In Section \ref{sec3}, we describe the objective and the scope of our study. The methodologies are thoroughly discussed. \\

\noindent (d) In Sections \ref{sec4}, the methodologies are implement with associated R codes. The mis-classification errors are discussed and recommendations are given.\\

\noindent (e) The conclusions and final recommendations are given in \ref{sec5}.\\

\Ni Finally, we provide our innovative R codes in the appendix after the bibliography.

\section{Description of the variables} \label{sec2}

\noindent Individual performance grants the student admission to Senior Secondary Education. The grading system in this result is categorized as follows : Credit (grades from 1 to 6), Pass (7 to 8), Fail (9). The result of a student is determined by the grades of the four core subject plus two best grades of other subjects choices. The best result a student could have is aggregate six, 1 in all core subjects plus 1 in any of the other subjects choices, and the worst result is aggregated 54 with failure  in all subjects with grade 9.\\

\noindent In this study, we focus on the core grades (G1), (G2), (G3) and (G4). Here are the conventions we will be using to know whether we use a grade for the whole population or not, for all the years or not, for both genders or not, or whether we use a grade for one particular gender, or a year or an area.\\

\noindent The corresponding variables are given as $G1$, $G2$, $G3$ and $G4$ before the cleaning and cover all areas, all time and all gender.\\

\noindent We also have the variables $year$ (from 2012 to 2017), gender (coded 1 for female and 2 for males), and the six regions (coded for 1 to 6).\\

\noindent Missing data (absences or submitted blank sheets) concern at least some cases. So we drop all cases presenting missing data. The new variables become :\\

\noindent G1A,  G2A,  G3A, G34, yearA, genderA, regionA.\\

\noindent When we restrict our analysis on a specific year for all gender, the variables are

\noindent G1A, G2A, G3A, G34, yearA, genderA, regionA.\\

\noindent When we restrict our analysis on a specific year for all genders all regions, the variables are:\\

\noindent G1AY,  G2AY, G3AY, G34Y, yearAY, genderAY,  regionAY.\\

\noindent When we restrict our analysis on a specific region for all genders and years, the variables are

\noindent G1AR,  G2AR, G3AR, G34R, yearAR, genderAR,  regionAR.\\

\noindent When we restrict our analysis on a specific region \textbf{and} for a specific year, for all genders, the variables are : \\

\noindent G1AYR, G2AYR, G3AYR, G34YR, yearAYR, genderAYR, regionAYR

\noindent At each labs, the specified year and region are given by the variables \textit{yearOfStudy} and \textit{regionOfStudy}.\\

\noindent If we want to specify and fix a gender, we use the codes Variable[gender=="1"] for female and Variable[gender=="2"] for male. For example :

$$
G1AR[genderAR==1], \ G2AYR[genderAYR==2],  \ G3A[genderA==1],
$$

\noindent respectively contain the grades of English for the region determined by the value of \textit{regionOfStudy}, the grades of Mathematics for the year and the region determined by the values assigned to \textit{yearOfStudy} and to \textit{regionOfStudy} respectively.\\

\noindent The notation above are fixed for once and used in similar papers. More notation related transformed variables may come later.\\

\section{Methodology} \label{sec3}

\noindent The problem is quite similar to assigning values to missing data. So we  preliminarily use methods from missing data analysis. But here, the context of the data will play a major role and will lead us to adopt hybrid solutions. Let us describe some of those methods.\\

\noindent \textbf{A - Linear regression}. Let us denote by T1, T2, T3 and T4 standing for the four grades G1, G2, G3 and G4 at the level we use them (all year and/or all regions, by year, by region, by gender, etc.).\\

\Ni We randomly select  $p=0.75\%$ of the complete data approximately as a training data, the rest being used as the test data. We denote the grades corresponding to the training by $TR1$, $TR2$, $TR3$ and $TR4$, and by $TS1$, $TS2$, $TS3$, $TS4$ for the test data.\\

\Ni Basically, the linear regression method linearly fits $TR4$ in $TR1$, $TR2$ and $TR3$. The regression model

\begin{equation}
\widehat{TR4} = C + a_{tr1} TR1 + a_{tr2} TR2 + a_{tr3} TR3, \ \label{mod_10}
\end{equation}  

\noindent is applied to the test data. For each observation $ts4$ in the test data is compared to its estimated $\hat{ts4}$ according to the model \eqref{mod_10}.\\

\noindent (A1) First of all, before even trying to apply it, we have to be sure that the regression is good enough. We may check the R-adjusted, the global Fisher p-value, the individual student p-values of the real contribution of each of the three grades, the AIC and/or the BIC. Let just work with the R-adjusted and the Fisher p-value. At what level of the R-adjusted should be good enough. In such an environment of human behavior, where exact formulas cannot be expected, R-ajusted's around $70\%$ seem to me good enough for us.\\

\noindent (A2) Once this step is successfully done, we may choose choose two types of assignments.\\

\noindent (A21) We assign the numerical value $\hat{ts4}$ to $ts4$ and the error of the estimation is $\hat{ts4}-ts4$. The global error of our model will be the aggregation of the individual errors in the test data in the form of the Euclidean, Max or Manhattan distance. However, judging the smallness of that error is not easy in this case. Indeed, we may judge the error is small and have $\hat{ts4}<9$ for a passing student in test data, and we are obliged to declare the case as failing. The second type we are going to discuss below seems to be more appropriate.\\

\noindent (A22) We assign a status of failing or passing according the estimated error of $\hat{ts4}$ is less than $9$ or not. We compute the rate of the two types of mis-classifications :\\

\noindent (A221) declaring the case as failing while it has passed (on the test data), corresponding to $(\hat{ts4}\geq 9) \wedge (ts4 <9) $ or rate $R_1$;\\

\noindent (A222) declaring the case as passing while it has failed (on the test data), corresponding to $(\hat{ts4}< 9) \wedge (ts4 =9) $ or rate $R_2$.\\

\noindent We adopt the model if these rates are small enough. Some times you cannot have both small enough. If one of the two errors implies more damage, we my decide to put a threshold on it and control the other, which should not be too big.\\

\Ni Before a definitive conclusion, we have to eliminate the randomness due to one choice of the training data. We proceed to a number $B$ of repetitions and take the average of the mis-classifications errors (that will be stable for $B$ large enough, according to the law of large numbers).\\

\noindent \textbf{B - Classification Method}.\\

\noindent It is theoretically possible to proceed by the $K$-means method. The $k$-means method classifies the training data into a fixed and chosen number $K$. We suppose that we have a distance between cases. If the distance allows it, we suppose that each class of finite cases can be represented by a centroid, a barycenter for example.  The algorithm works as follows :\\

\begin{enumerate}
\item Take $K$, the $K$ individuals which are the most distant between them among all individuals. Create $K$ classes from each of them.
\item Repeat the following procedure until there exists no case to classify
	\begin{enumerate}
		\item For each case not already classified, find the class which is closest to it.
		\item Add it to that class
		\item compute the new centroid of the class 
	\end{enumerate}
\end{enumerate}  

\Bin Once we finish creating the $K$ classes, we set the following rule.\\

\Ni \textbf{Assignment of the missing grade}. For each case in the test data, we determine the closest class and assign to it either the average $T4$ grade over the class or the most frequent $T4$ grade. We compute the two types of errors as previously.\\

\noindent \textbf{C - Nearest-neighbor method}.\\

\Ni Here it works similarly as the method B, with the working distance using the grades $T1$, $T2$ and $T3$ - between cases. We proceed as follows:\\

\noindent (a1) We fix $K$ number of neighbors,\\
 
\noindent (b) For each case of the test data, we compute the distances of that case and all the cases of the training data. We set a class of neighbors by picking the $K$ closest cases of the training data.\\

\noindent (c) \textbf{Assignment of the missing grade}. We assign to the treated case either the average $T4$ grade over the class of neighbors or the most frequent $T4$ grade over the class of neighbors. We compute the two types of errors as previously.\\

\Ni Here we fixed the number of neighbors, But we might let it vary and instead, choose a threshold $\varepsilon$ and take the neighborhood as a ball of radius $\varepsilon$ and centered at the treated case.\\

\noindent (a2) The neighbors are all cases in the training data not far from the treated case more than $\varepsilon$.\\

\noindent (c2) \textbf{Assignment of the missing grade}. We assign to the treated case either the average $T4$ grade over neighborhood  or the most frequent $T4$ grade over neighborhood . We compute the two types of errors as previously.\\

\section{Implementation of the methods and discussion of the results} \label{sec4}

\noindent \textbf{A - Non favorable conditions of application of each method}.\\

\noindent The data present a great number of repetitions. While the methods are based on classifications of distances, the data seem to be regrouped in class of inter-distance equal to zero.\\

\noindent This fact automatically excludes the $K$-means method since for $K$ fixed, we may have that the class of maximum inter-distance has more that $K$ elements. How could we pick $K$ of them to form the original classes. Furthermore, even if we pick $K$ of them randomly, classifying the other cases is difficult since a non classified case can have the same distance with all of the $K$ classes. Hence, we dismiss the method.\\

\noindent By implementing the KNN method, we also have the same problem of choosing $K$ neighbors, since they may have far more that $K$ elements in the training data whose distance to the treated case is zero.\\

\Ni From this set of remark, we propose a mixture of the classification and the KNN methods as follows.

\noindent \textbf{B - A hybrid method}. We propose a follows.\\

\noindent (a) Fix a number $K$ of neighbors\\

\noindent (b) Find the number $KT$ of cases in the training data that are equal to the treated cases.\\

\noindent (b1) If $KT\geq K$, we take all the cases with zero distance to the treated case and we use it as the assignment class.\\

\noindent (b2) If $KT<K$, we form the assignment class by taking picking all the cases with zero distance to the treated case, of number $KT$ and we complete them to $K$ by using the $K-KT$ closest cases form the remaining cases of the training data.\\

\Ni We are going to implement that hybrid method and see if it gives a solid model.\\

\Ni \textbf{B - Local approach}. We suggest to use a local approach in the following sense. If we treat  a particular case, it make sense to think that the model should be better if we use similar cases to it, meaning we should remain in the year and in the same region. We will see if the gender also might be influential.\\

\Ni So, theoretically, we may try to build models for any year and any region that will give 36 models. Instead, we will focus on models which will be effectively used. Indeed from the current data, we may find all the cases of individuals who missed grade G4 and passed the other grades. We already listed them in Table \ref{tab1}.\\

\noindent To find the contents of that table, we run the codes (LRCAD) on page \pageref{LRCAD}. Related explanations are given there. So, we are going to build the models for the year 2015 and regions 1,2 and 4, and for year 2017 and region 1. We will fully explain the methodologies for one choice (year and region). For the other choices, we simply give the outputs and their analysis.\\

\Ni Let us begin with the analysis for year 2015 and region 1.\\

\Ni \textbf{C - Building models}.\\

\Ni \textbf{C1 - Building models for (2015, 1)}.\\

\Ni \textbf{C1A - The regression models}. Let us discover the outputs after running the algorithm (LMYR), page \pageref{LMYR}.\\

\Ni \textbf{I - The outcomes}. For each pair of year of region as given in Table \ref{tab2}, we run the codes $B=100$ times. In the left table, we display the R adjusted coefficients of the linear regressions. In the right table, we display the average of two kinds of errors \textit{(mpf, \ mfp)} where :\\

\noindent (a) \textit{mpf} is the rate of failing knowing that the case passed;

\noindent (b) \textit{mfp} is the rate of passing knowing that the case failed.  

\begin{table}[h]
		\begin{tabular}{|lll|}
		\hline \hline
			& 2015 	& 	2017\\
		\hline \hline
		1 & 78\%			& 74\%\\
		2 & 78\%			& \\
		3 &  74\%		&\\
		\hline \hline
		\end{tabular}
		\begin{tabular}{|lll|}
			\hline \hline
			& 2015 							& 2017\\
			\hline \hline
		1 & (16.99 \%, 0.30\%)			& (16.38\%, 0.44\%)\\
		2 &  (25.10\%, 0.15\%)			& \\
		3 & (8.66 \%, 0.24\%)				& \\
		\hline \hline
		\end{tabular}
		\caption{Tables of R-adjusted vales (left), pair of mis-classification erreors (right)}
		\label{tab2}
\end{table}

\Ni \textbf{II - Evaluating the model}.\\

\Ni The model is relatively good. From the current data, we have the following overall facts.\\

\Ni \textbf{First error}. The probability of a failing student to be pass through our method is very weak, less that $1\%$.

\Ni \textbf{Second error}. The probability of a passing student to be fail through our method is around $20 \%$. Sometimes, it much weaker.\\

\Ni How to interpret the results? A student $S$ has missed the exam of grade $G4$. \textit{Normally}, $S$ has failed according to the general rule that one cannot miss an exam and pass. Now $S$ is given a second chance by our rule. The research says that :\\

\Ni If $S$ were a failing student, he/she wouldn't be passing in the actual rule. In that sense, the rule is fair since it would be give an advantage to students who missed the exam. If the contrary would have happened, it would give an excuse to students and let them say : \textit{let us miss the exam. The statistical system will assign us undue chances to pass}.\\

\Ni If $S$ were a passing, there is non-negligible chance of failing through the statistical system of around $20\%$, But given that the first outcome is that he/she should fail, but now he/she still has $80\%$ chances to recover a passing grade. Although he/she does not have $100\%$ chance to recover his/her passing status, he/she is now lucky for not failing with $80\%$ chance.\\

\Ni We consider that the statistical system is fair and useful to unlucky good students who unintentionally missed the grade four exam.\\

\Ni \textbf{III - Application}.\\

\Ni Let us apply the the rule to the candidates in Table \ref{tab1}. The general methodology is the following.\\

\Ni \textbf{1.} We fix the case (\textit{caseT}) to be predicted as VC=c(G1[caseT], G2[caseT], G3[caseT], G4[caseT]). Of course, we should check that $G4[caseT]$ is missing, that is $G4[caseT]="-1"$.\\

\noindent \textbf{2.} To be in similar conditions as in the learning process, we randomly select a set of $75\%$ cases in the complete data. The selected data are $TT1$ (extracted from $G1$), $TT2$ (from $G2$) and $TT3$ (from $G3$) and $TT4$ (from $G4$). We apply the linear model 

$$
lm1 = lm(TT4 ~ TT1 + TT2 + TT3)
$$

\noindent The coefficients of the model are given by 

\begin{lstlisting}
lm1$coef[1] (constant), lm1$coef[2] (coefficient of TT1),
 lm1$coef[3] (coefficient of TT2), lm1$coef[4] (coefficient of TT4).
\end{lstlisting}

\noindent The predicted grade is 

\begin{lstlisting}
VCE=lm1$coef[1]+(lm1$coef[2]*VC[1])+(lm1$coef[2]*VC[2])
+(lm1$coef[3]*VC[3]).
\end{lstlisting}

\Ni For $B=100$ random choices of the training data, we compute the rate of passing on the overall regressions using the codes. Before the loop on $B$, we set
$grade4F=0$, $grade4P=0$. At each passage, we count the successes by using

\begin{lstlisting}
	if(VCE>8){
		grade4F=grade4F+(1/B)
	}
	else{
	grade4P=grade4P+(1/B)
	}
\end{lstlisting}

\Ni After the $B$ passage, we grant a passing grade if grade4P exceeds $50\%$. Otherwise, we let it go as failing.\\
 
\Ni Here are our predictions of the valid cases to be treated.\\

\Ni \textbf{Case caseT=80 183 (Year 2015, Region 1)}.  $grade4P=100\%$. Pass granted.\\

\Ni \textbf{Case 77 594 (Year 2015, Region 2)}.  $grade4P=100\%$. Pass granted.\\

\Ni \textbf{Case 77 833 (Year 2015, Region 3)}.  $grade4P=100\%$. Pass granted.\\

\Ni \textbf{Case 122 915 (Year 2017, Region 1)}.  $grade4P=100\%$. Pass granted.\\

\newpage

\Ni \textbf{C1B - The hybrid Classification-KNN}. Let us discover the outputs after running the algorithm (HCN), page \pageref{HCN}.\\

\Ni \textbf{I - The models}.\\

\Ni Here we proceed in two stages. First of all we fix $K=100$ the number of neighbors we use for our estimation. We randomly choose 75\% (approximately) of the complete data as training data. For each each case of the test data, we find the number $K_{sim}$ of all \textit{similar} cases with respect to the first three grades. We consider two models :\\

\Ni \textbf{Model 1}. If $K_{sim}>K$, we take the class $\mathcal{C}_{sim}$ of all similar cases for deciding. Here again, we consider two approaches.

\Ni \textbf{Model 1a}. We compute the average $TA4$ of the fourth grade $T4$ over the class $\mathcal{C}_{sim}$ and grant a pass grade  if $TA4\leq 8$ or a fail grade for $TA4>8$. The two errors are denoted \textit{mpfSIMAVM1} (for passing but failing in the rule) and \textit{mfpSIMAVM1} (for falling but passing in the rule).\\

\Ni \textbf{Model 1b}. We take the most frequent result $TMF4$ for the fourth grade $T4$ over the class $\mathcal{C}_{sim}$ and grant a pass grade  if $TMF4\leq 8$ or a fail grade for $TA4>8$. The two errors are denoted \textit{mpfSIMMFM2} (for passing but failing in the rule) and \textit{mfpSIMMFM2} (for falling but passing in the rule).\\

\Ni Let us just help on how to remember the type of error with \textit{mpfSIMAVM1}, that we can write as mpf-SIM-AV-M1, meaning : \textit{mpf} (\textbf{m}isclassification for \textbf{p}assing case that \textbf{f}ails in the statistical rule), \textit{SIM} (when we take all the \textbf{sim}ilar cases 
as a basis of decision), \textit{AV} (we use the \textit{av}erage to grant pass or not), \textit{M1} (in Model 1). Below, we may replace \textit{AV} by \textit{MF} when we use the most frequent grade instead of the average. We may also replace \textit{SIM} by \textit{NN} whn we use the nearest neighbors, etc..\\

\Ni \textbf{Model 2}. If $K_{sim}<K$. We form the estimating class $\mathcal{C}_{ng}$ by including first all the similar case. Next, we complete $\mathcal{C}_{ng}$ by taking the $(K-K_{sim})$ most close cases, where we use the euclidean distance. Because of the nature of the data, it is possible that we will not be able to have $(K-K_{sim})$ closest neighbors \textit{exactly}, since non-chosen cases may be similar to chosen cases. However, in such a situation, we only use some of them and disregard the others. As in Model 1, we proceed with two approaches.\\

\Ni \textbf{Model 2a}. We compute the average $TA4$ of the fourth grade $T4$ over the class $\mathcal{C}_{ng}$ and grant a pass grade  if $TA4\leq 8$ or a fail grade for $TA4>8$. The two errors are denoted \textit{mpfNNAVM2} (for passing but failing in the rule) and \textit{mfpNNAVM2} (for falling but passing in the rule).\\

\Ni \textbf{Model 2b}. We take the most frequent result $TMF4$ for the fourth grade $T4$ over the class $\mathcal{C}_{ng}$ and grant a pass grade  if $TMF4\leq 8$ or a fail grade for $TA4>8$. The two errors are denoted \textit{mpfNNMFM2} (for passing but failing in the rule) and \textit{mfpNNMFM2} (for falling but passing in the rule).\\

\Ni After we run the algorithm (HCN) a number of $B=10$ to $B=100$ times, the good performances we have are the following.\\

\Ni \textbf{Performance by the similar cases Method}

\begin{table}[h]
	\centering
	\begin{tabular}{ccc}
	\hline \hline
	\multicolumn{3}{c}{Model 1}\\
				&  Model 1a (average) & Model 1b (most frequent)\\
	\hline \hline
	mpf		&	 10.5\%					&	20.6\%			\\
	mfp		&	 79.4\%					&	0\%			\\
\hline \hline
\end{tabular}
\end{table}

\Ni \textbf{Recommendation}. We recommend use of Model 1b for estimating the grade for a case whose number of similar cases is greater than the number of neighbors $K$. The model consists of assigning the most frequent grade as an estimate of the missing grade.

\Ni \textbf{Performance by the nearest neighbor cases Method}

\begin{table}[h]
	\centering
\begin{tabular}{ccc}
\hline \hline
\multicolumn{3}{c}{Model 2}\\
\hline \hline

				&  Model 2a (average) & Model 2b (most frequent)\\
\hline \hline
	mpf		&	 20,6\%					&	88.9\%			\\
	mfp		&	 11.11\%					&	0\%			\\
\hline \hline
\end{tabular}
	
\end{table}

\Ni \textbf{Recommendation}. We recommend use of Model 2a for estimating the grade of a case whose number of similar cases is less than the number of neighbors $K$. The model consists of taking the average grade as an estimate of the missing grade.

\Ni \textbf{II - Predictions}.\\

\Ni Here are our predictions of the valid cases to be treated.\\

\Ni \textbf{Case caseT=80 183 (Year 2015, Region 1)}.  The KNN method is used. Mean grade : $4.63$. Most frequent grade in the neighbors : 3. Pass granted.\\

\Ni \textbf{Case 77 594 (Year 2015, Region 2)}.  The KNN method is used. Mean grade : $7.4$. Pass granted by the rule of Model 2a.\\

\Ni \textbf{Case 77 833 (Year 2015, Region 3)}. The KNN method is used. Mean grade : $7.31$. Pass granted by the rule of Model 2a.\\ 

\Ni \textbf{Case 122 915 (Year 2017, Region 1)}. The KNN method is used. Mean grade : $4.61$. Most frequent grade in the neighbors : 5. Pass granted.\\

\textbf{Remark}. For cases $77 594$ and $77 833$, the model 2b is not stable. The failing grade is the most frequent grade. But we already ruled that the model 2a is more accurate. 

\section{Conclusion} \label{sec5}

\Ni we can say that with a controllable errors of mis-classifications, we succeeded in setting a statistical rule that can assign a passing grade to students who missed the fourth exam unintentionally. We proposed two methods: the regression method, and a hybrid method using a combination of classification and nearest neighbors.\\

\Ni Both rules are fair in the sense that really failing students have a very few chance to pass through the statistical device. They allow to deserving students to be granted a passing grade with around 20\% of spoiling their chance.\\

\Ni Study focused on estimating missing SES grades. But the method and the codes can easily be adapted with a simple permutation. But we recommend to run our R codes in each case to proceed as follows.\\

\noindent (a) Choose the grade to be estimated.\\

\noindent (b) Run all the script \textit{2019\_11\_26\_file\_01\_data\_and\_functions.R} which will read the data. That script contains all the functions we need.\\

\noindent (c) To apply the regression method,

\begin{itemize}
\item[\textbf{(c2)}] Run the script \textit{2019\_11\_26\_file\_02\_models\_regressions.R} to build the model. Interpret the mis-classifications errors and decide if the method should be used or not.
\item[\textbf{(c1)}] To predict a missing case, run the script \textit{2019\_11\_26\_file\_03\_models\_regressions\_prediction.R} in which the existed grades should be explicitly given in the first line.
\end{itemize}

\noindent (d) To apply the \textit{HVN} method,

\begin{itemize}
\item[\textbf{(d1)}] Run the script \textit{2019\_11\_26\_file\_04\_models\_hybrid\_Classi\_NN.R} to build the model. Interpret the mis-classifications errors and decide if the method should be used or not 
\item[\textbf{(d2)}] To predict a missing case, run the script \textit{2019\_11\_26\_file\_05\_models\_hybrid\_Classi\_NN\_prediction\_ok.R} in which the existed grades should be explicitly given in the first line.
\end{itemize}

\newpage
\normalsize

\section{Appendix : R codes} \label{sec5}

\Ni The R codes for reading the data are given in \cite{saine2020}. The software R (see \cite{r2020}) is used. For a wide overview of R, see \cite{crawley}.\\

\Ni \textbf{1. List of rescuable cases over all the data (LRCAD)}. \label{LRCAD}

\begin{lstlisting} 
10 T1=G1; T2=G2; T3=G3; T4=G4; 
11 N=length(T1)

12 RESCUYR <- numeric(); RESCU <- numeric()
13 h=1; r=4; s=0

14	for(h in 1:N){
			vp=c(T1[h], T2[h], T3[h], T4[h])
			if(rescuable(vp,r)==1){
				s=s+1
				RESCUYR[s]=h
		}
	}

22 NRESCUYR=s

23 #edit(cbind(T1[RESCUYR], T2[RESCUYR], T3[RESCUYR],T4[RESCUYR]))

T1S=T1[RESCUYR]; T2S=T2[RESCUYR]; T3S=T3[RESCUYR]; T4S=T4[RESCUYR]
T1SC=T1S[(T1S<9)&(T2S<9)&(T3S<9)]
T2SC=T2S[(T1S<9)&(T2S<9)&(T3S<9)]
T3SC=T3S[(T1S<9)&(T2S<9)&(T3S<9)]
T4SC=T4S[(T1S<9)&(T2S<9)&(T3S<9)]

29 #edit(cbind(T1SC, T2SC, T3SC, T4SC))
30 index=seq(1,N)

V1=T1[(T1>0) & (T1<9) & (T2>0) & (T2<9) & (T3>0) & (T3<9) & (T4==-1)]
V2=T2[(T1>0) & (T1<9) & (T2>0) & (T2<9) & (T3>0) & (T3<9) & (T4==-1)]
V3=T3[(T1>0) & (T1<9) & (T2>0) & (T2<9) & (T3>0) & (T3<9) & (T4==-1)]
V4=T4[(T1>0) & (T1<9) & (T2>0) & (T2<9) & (T3>0) & (T3<9) & (T4==-1)]
Y1=year[(T1>0) & (T1<9) & (T2>0) & (T2<9) & (T3>0) & (T3<9) & (T4==-1)]
R1=region[(T1>0) & (T1<9) & (T2>0) & (T2<9) & (T3>0) & (T3<9) & (T4==-1)]
index1=index[(T1>0) & (T1<9) & (T2>0) & (T2<9) & (T3>0) & (T3<9) & (T4==-1)]

38 #edit(cbind(index1,Y1,R1,V1,V2,V3,V4))
\end{lstlisting}

\newpage
\Ni \textbf{2. Global package to be run for once  (LMYR)}. \label{LMYR}\\

\Ni \textbf{First Part I : choose the region and the year}.\\

\begin{lstlisting}
yearOfStudy=2015
regionOfStudy=2
genderY=gender[year==yearOfStudy]
regionY=regionA[year==yearOfStudy]
G1Y=G1[year==yearOfStudy]
G2Y=G2[year==yearOfStudy]
G3Y=G3[year==yearOfStudy]
G4Y=G4[year==yearOfStudy]
#========================
genderR=gender[region==regionOfStudy]
G1R=G1[region==regionOfStudy]
G2R=G2[region==regionOfStudy]
G3R=G3[region==regionOfStudy]
G4R=G4[region==regionOfStudy]

# study of region and year

genderYR=gender[(region==regionOfStudy) & (year==yearOfStudy)]
G1YR=G1[(region==regionOfStudy) & (year==yearOfStudy)]
G2YR=G2[(region==regionOfStudy) & (year==yearOfStudy)]
G3YR=G3[(region==regionOfStudy) & (year==yearOfStudy)]
G4YR=G4[(region==regionOfStudy) & (year==yearOfStudy)]
#length(G4YR)

#c=c(1,2,3,4) # exclut G4
#c=c(1,2,4,3) # exclut G3
#c=c(1,3,4,2) # exclut G2
#c=c(2,4,3,1) # exclut G1

#T1=G1YR; T2=G2YR; T3=G3YR; T4=G4YR;
g=2
T1=G1YR[genderYR==g]; T2=G2YR[genderYR==g]; 
T3=G3YR[genderYR==g]; T4=G4YR[genderYR==g];
#T1=G1YR; T2=G2YR; T3=G3YR; T4=G4YR;
\end{lstlisting}

\newpage

\Ni \textbf{Part II : Run the loo[}

\begin{lstlisting}
b=1
B=100
mpfB=0
mfpB=0
#B=1
for(b in 1:B){

#length(T1)

#####  Training - testing Data

(N=length(T1))
(n=ceiling((2*N)/3))
(n=1.5*round((2*N)/3)+ 2000)
(t=round(runif(n,1,N)))

td<-numeric()
k=1
td[1] = t[1]
for(h in 2:n){
   if(presenceTest(t[h],td,k)==0){
     k=k+1
     td[k] = t[h]
   }
}
#sort(td)
N=length(T1)
n=k
(100*n/N)


# Supervised Learning.
#Traing Data.

(TT1=T1[td])
(TT2=T2[td])
(TT3=T3[td])
(TT4=T4[td])

length(TT1)

#Test data

(TS1=T1[-td])
(TS2=T2[-td])
(TS3=T3[-td])
(TS4=T4[-td])
#length(TS1)
#length(TT1)+length(TS)
(ntrain=n)
(ntest=N-n)

lm1 = lm(TT4 ~ TT1 + TT2 + TT3)
#summary(lm1)


TSE=lm1$coef[1]+(lm1$coef[2]*TS1)+(lm1$coef[2]*TS2)+(lm1$coef[3]*TS3)

#### Errors

mpf=0
mfp=0
totalpass=0
totalfail=0
h=0

for(h in 1:ntest){

	#if((TS4[h]<9) & (TS1[h]<9) & (TS2[h]<9) & (TS3[h]<9) ){
	if((TS4[h]<9) ){
			totalpass=totalpass+1
		
			if(TSE[h]>8){
				mpf=mpf+1
			}

	}
	else{
		totalfail=totalfail+1
		if(TSE[h]<0){
			mfp=mfp+1
		}

	}
}
#totalpass
#totalfail
#totalpass+totalfail
mpfB=mpfB+ ((mpf/totalpass)/B)
mfpB=mfpB+ ((mfp/totalfail)/B)
}
summary(lm1)
100*mpfB
100*mfpB
\end{lstlisting}

\newpage
\Ni \textbf{3. Prediction for new cases (RNC)}. \label{RNC}\\

\begin{lstlisting}
#caseT=80183
#caseT=77833
#caseT=77594
caseT=122915
VC=c(G1[caseT], G2[caseT], G3[caseT], G4[caseT])
#VC
###############################
yearOfStudy=2017
regionOfStudy=1

genderY=gender[year==yearOfStudy]
regionY=regionA[year==yearOfStudy]
G1Y=G1[year==yearOfStudy]
G2Y=G2[year==yearOfStudy]
G3Y=G3[year==yearOfStudy]
G4Y=G4[year==yearOfStudy]
#========================
genderR=gender[region==regionOfStudy]
G1R=G1[region==regionOfStudy]
G2R=G2[region==regionOfStudy]
G3R=G3[region==regionOfStudy]
G4R=G4[region==regionOfStudy]

# study of region and year
indexC=seq(1,length(G1))
length(indexC)
genderYR=gender[(region==regionOfStudy) & (year==yearOfStudy)]
G1YR=G1[(region==regionOfStudy) & (year==yearOfStudy)]
G2YR=G2[(region==regionOfStudy) & (year==yearOfStudy)]
G3YR=G3[(region==regionOfStudy) & (year==yearOfStudy)]
G4YR=G4[(region==regionOfStudy) & (year==yearOfStudy)]
indexCYR=indexC[(region==regionOfStudy) & (year==yearOfStudy)]
#length(G4YR)

#c=c(1,2,3,4) # exclut G4
#c=c(1,2,4,3) # exclut G3
#c=c(1,3,4,2) # exclut G2
#c=c(2,4,3,1) # exclut G1

#T1=G1YR; T2=G2YR; T3=G3YR; T4=G4YR;
g=2
T1=G1YR[genderYR==g]; T2=G2YR[genderYR==g]; 
T3=G3YR[genderYR==g]; T4=G4YR[genderYR==g];

T1=G1YR; T2=G2YR; T3=G3YR; T4=G4YR;
b=1
B=10
mpfB=0
mfpB=0
#B=1
grade4P=0
grade4F=0

for(b in 1:B){
#####  Training - testing Data
(N=length(T1))
(n=ceiling((2*N)/3))
(n=1.5*round((2*N)/3)+ 2000)
(t=round(runif(n,1,N)))

td<-numeric()
k=1
td[1] = t[1]
for(h in 2:n){
   if(presenceTest(t[h],td,k)==0){
     k=k+1
     td[k] = t[h]
   }
}
#sort(td)
N=length(T1)
n=k
(100*n/N)

# Supervised Learning.
#Traing Data.
(TT1=T1[td])
(TT2=T2[td])
(TT3=T3[td])
(TT4=T4[td])
#length(TT1)

#Test data
(TS1=T1[-td])
(TS2=T2[-td])
(TS3=T3[-td])
(TS4=T4[-td])
#length(TS1)
#length(TT1)+length(TS)
(ntrain=n)
(ntest=N-n)

lm1 = lm(TT4 ~ TT1 + TT2 + TT3)
#summary(lm1)

VCE=lm1$coef[1]+(lm1$coef[2]*VC[1])+(lm1$coef[2]**VC[2])+(lm1$coef[3]**VC[3])
	if(VCE>8){
		grade4F=grade4F+(1/B)
	}
	else{
	grade4P=grade4P+(1/B)
	}
}
grade4P
\end{lstlisting}

\newpage
\Ni \textbf{4. Building models Hybrid Classification-KNN (HCN)}. \label{HCN}\\ 

\Ni (Needs the functions : presencetest, distance, upperPartialmean, lowerPartialMean below).

\begin{lstlisting}
yearOfStudy=2015
regionOfStudy=1


genderAY=genderA[yearA==yearOfStudy]
#length(genderAY)

regionAY=regionA[yearA==yearOfStudy]
#length(regionAY)

G1AY=G1A[yearA==yearOfStudy]
length(G1AY)

G2AY=G2A[yearA==yearOfStudy]
length(G2AY)

G3AY=G3A[yearA==yearOfStudy]
#length(G3AY)

G4AY=G4A[yearA==yearOfStudy]
#length(G4AY)

# study of region
genderAR=genderA[regionA==regionOfStudy]
G1AR=G1A[regionA==regionOfStudy]
G2AR=G2A[regionA==regionOfStudy]
G3AR=G3A[regionA==regionOfStudy]
G4AR=G4A[regionA==regionOfStudy]

# study of region and year

genderAYR=genderA[(regionA==regionOfStudy) & (yearA==yearOfStudy)]
G1AYR=G1A[(regionA==regionOfStudy) & (yearA==yearOfStudy)]
G2AYR=G2A[(regionA==regionOfStudy) & (yearA==yearOfStudy)]
G3AYR=G3A[(regionA==regionOfStudy) & (yearA==yearOfStudy)]
G4AYR=G4A[(regionA==regionOfStudy) & (yearA==yearOfStudy)]
#length(G4AYR)

#proceed to your analysis
# Choose type of data
W1 <- numeric()
W2 <- numeric()
W3 <- numeric()
W4 <- numeric()
V1 <- numeric()
V2 <- numeric()
V3 <- numeric()
V4 <- numeric()

W1=G1AYR
W2=G2AYR
W3=G3AYR
W4=G4AYR
(NT=length(W4))
#transform values credit 1 between 1 and 6, pass between 7 and 8, 3 fail for 9
for(j in 1:NT){
	if(W1[j]<7){
		V1[j]=1
	}
	else{
		if(W1[j]<9){
			V1[j]=2
		}
		else{
			V1[j]=3
		}
	}

	if(W2[j]<7){
		V2[j]=1
	}
	else{
		if(W2[j]<9){
			V2[j]=2
		}
		else{
			V2[j]=3
		}
	}

	if(W3[j]<7){
		V3[j]=1
	}
	else{
		if(W3[j]<9){
			V3[j]=2
		}
		else{
			V3[j]=3
		}
	}

	if(W4[j]<7){
		V4[j]=1
	}
	else{
		if(W4[j]<9){
			V4[j]=2
		}
		else{
			V4[j]=3
		}
	}
}
V1
V2
V3
V4
##   Choice of de X1, X2, X3, X4
X1=W1;X2=W2; X3=W3; X4=W4
#X1=V1;X2=V2; X3=V3; X4=V4
## ================================
## BIGGEST LOOP FOR B EXPERIENCES

B=100
b=1
mpfMG1=0
mfpMG1=0
mpf2MG1=0
mfp2MG1=0

mpfMG2=0
mfpMG2=0
mpf2MG2=0
mfp2MG2=0

for(b in 1:B){
# grande boucle

(N=length(X4))
(n=ceiling((2*N)/3))
(n=1.5*round((2*N)/3)+ 2000)
(t=round(runif(n,1,N)))

#length(X4)
#n

td<-numeric()
k=1
td[1] = t[1]
for(h in 2:n){
   if(presenceTest(t[h],td,k)==0){
     k=k+1
     td[k] = t[h]
   }
}
#sort(td)
k
(n=k)
(100*n/N)


# Supervised Learning.
#Traing Data.

(XT1=X1[td])
(XT2=X2[td])
(XT3=X3[td])
(XT4=X4[td])

length(XT1)

#Test data
(XTD1=X1[-td])
(XTD2=X2[-td])
(XTD3=X3[-td])
(XTD4=X4[-td])
length(XTD1)
length(XT1)+length(XTD1)
ntrain=n
ntest=N-n

#Building the model
#K=20
# Estimation of X4 for the first testing data.
c<- numeric()
p<- numeric()
D<-numeric()
freq<- numeric()
freqMFM <- numeric()
distTestToTrain <- numeric()
group<- numeric()
F<- numeric()
h=1
size = 3
q=0
o=1
gradex4=0
sum=0
v=1
K=100 #nearest neighbors

for(s in 1:ntest){
	for(u in 1:9){freq[u]=0}
  	q=0
  	gradex4=0
	c=c(XTD1[s], XTD2[s], XTD3[s])

	for(r in 1:ntrain){
 
  	p=c(XT1[r], XT2[r], XT3[r])
  	sum=0
	distTestToTrain[r]=distance(h,p,c,size)
  	for(o in 1:size){
    	if(p[o] != c[o]) sum=sum+1
   	} 
		if(sum==0){
   		q=q+1 
   		gradex4=gradex4+XT4[r]
		for(u in 1:9){
			if(XT4[r]==u) freq[u]=freq[u]+1
		}
 	}
	
}
	if(q>K){
		group[s]=1
		D[s] = gradex4/q
		F[s]=freq[order(freq)][9]
	}
	else{
		group[s]=-1
		D[s]=lowerPartialMean(XT4[order(distTestToTrain)],K)
		F[s]=0

			for(u in 1:7){freqMFM[u]=0}
			for(v in 1:K){
				for(u in 1:9){
					if(XT4[order(distTestToTrain)][v]==u){
						freqMFM[u]=freqMFM[u]+1
					}
				}
			}
		F[s]=freqMFM[order(freqMFM)][9]


	}
}

# misclassification
g=1
Z1=XTD4[group==g]
Z2=D[group==g]
Z3=F[group==1]

#length(Z1)
#length(Z3)
nsg=length(Z1)
mpf=0
mfp=0
mpf2=0
mfp2=0

for(r in 1:nsg){
		if(Z1[r]<9){
			if(Z2[r]>8){
				mpf=mpf+1
			}

			if(Z3[r]>8){
				mpf2=mpf2+1
			}
		}
		else{
			if(Z2[r]<9){
				mfp=mfp+1
			}

			if(Z3[r]<9){
				mfp2=mfp2+1
			}
		}
}

mpfMG1=mpfMG1+ ((mpf/nsg)/B)
mfpMG1=mfpMG1+ ((mfp/nsg)/B)
mpf2MG1=mpf2MG1+ ((mpf2/nsg)/B)
mfp2MG1=mfp2MG1+ ((mfp2/nsg)/B)

g=-1
Z1=XTD4[group==g]
Z2=D[group==g]
Z3=F[group==1]

#length(Z1)
#length(Z3)
nsg=length(Z1)
mpf=0
mfp=0
mpf2=0
mfp2=0

for(r in 1:nsg){
		if(Z1[r]<9){
			if(Z2[r]>8){
				mpf=mpf+1
			}

			if(Z3[r]>8){
				mpf2=mpf2+1
			}
		}
		else{
			if(Z2[r]<9){
				mfp=mfp+1
			}

			if(Z3[r]<9){
				mfp2=mfp2+1
			}
		}
}

mpfMG2=mpfMG2+ ((mpf/nsg)/B)
mfpMG2=mfpMG2+ ((mfp/nsg)/B)
mpf2MG2=mpf2MG2+ ((mpf2/nsg)/B)
mfp2MG2=mfp2MG2+ ((mfp2/nsg)/B)
# end of the big loop
}

# Group +1
100*mpfMG1
100*mfpMG1
100*mpf2MG1
100*mfp2MG1

# Group -1
100*mpfMG2
100*mfpMG2
100*mpf2MG2
100*mfp2MG2
\end{lstlisting}


\newpage
\Ni \textbf{Functions}.\\


\Ni \textbf{1. Function presence of an element in a current array}

\begin{lstlisting}
#Function Presence Test
L=c(0,2,5,8)
freq <- numeric()
test=0
xx=1
NN=1

presenceTest <- function(xx,L,NN){
  test=0
  for(h in 1:(NN)){
    if(L[h]==xx){
      test=1
    }
  }
  return(test)
}

# End of function Presence Test
\end{lstlisting}

\Ni \textbf{2. Average of lower elements}

\begin{lstlisting}
c<- numeric()
p <- numeric()
size=1
dist=0
average=0
r=0
subsize=1
totsize=1

lowerPartialMean <- function(c,size){
	average=0
	for(r in 1:size){
			average=average+c[r]
	}
	average=average/size
	return(average)
}
\end{lstlisting}

\Ni \textbf{3. Average of lower elements}

\begin{lstlisting}
upperPartialMean <- function(c,subsize,totsize){
	average=0
	for(r in 1:subsize){
			average=average+c[totsize-r+1]
	}
	average=average/subsize
	return(average)
}

c=c(1,0,3,6,10)
upperPartialMean(c,2,5)
c[4], c[5]
\end{lstlisting}

\Ni \textbf{4. Distance between two vectors}

\begin{lstlisting}
distance <- function(h,p,c,size){
  if(h==1){
    dist=0
    for(j in 1:size){
      dist=dist+(p[j]-c[j])^2
    }
    return(dist)
  }
  
  if(h==2){
    dist=0
    for(j in 1:size){
      if(abs(p[j]-c[j])>dist){
        dist=abs(p[j]-c[j])
      }
    }
    return(dist)
  }
}
#End of distance function
\end{lstlisting}

\end{document}